\documentclass[fleqn,usenatbib]{mnras}

\usepackage{newtxtext,newtxmath}
\usepackage{xcolor}     
\definecolor{textRed}{rgb}{0.00, 0.00, 0.00}
\newcommand{\textRed}[1]{\textcolor{textRed}{#1}}

\usepackage[T1]{fontenc}

\DeclareRobustCommand{\VAN}[3]{#2}
\let\VANthebibliography\thebibliography
\def\thebibliography{\DeclareRobustCommand{\VAN}[3]{##3}\VANthebibliography}

\usepackage{graphicx}	
\usepackage{amsmath} 
\usepackage{bm}

\usepackage[utf8x]{inputenc}

\title[BOSS LRG Marked correlation function]{A new test of gravity - II: \textRed{Application of marked correlation functions to luminous red galaxy samples}}

\author[J. Armijo et al.]{
Joaquin Armijo,$^{1,2}$\thanks{E-mail: joaquin.a.armijo-torres@durham.ac.uk}
Carlton M. Baugh,$^{2}$
Peder Norberg$^{2,3}$
Nelson D. Padilla$^{4}$
\\
$^{1}$Center for Data-Driven Discovery, Kavli IPMU (WPI), UTIAS, The University of Tokyo, Chiba 277-8583, Japan\\
$^{2}$Institute for Computational Cosmology, Department of Physics, Durham University, South Road, Durham, DH1 3LE, UK\\
$^{3}$Centre for Extragalactic Astronomy, Department of Physics, Durham University, South Road, Durham DH1 3LE, UK\\
$^{4}$Instituto de Astromía Teórica y Experimental (IATE), CONICET
Universidad Nacional de Córdoba, Laprida 854, X5000BGR, Córdoba, Argentina
}

\date{Accepted XXX. Received YYY; in original form ZZZ}

\pubyear{2022}

\begin{document}
\label{firstpage}
\pagerange{\pageref{firstpage}--\pageref{lastpage}}
\maketitle

\begin{abstract}
We apply the marked correlation function test proposed by Armijo et al. (Paper I) to samples of luminous red galaxies (LRGs) from the final data release of the Sloan Digital Sky Survey (SDSS) III. The test assigns a \textRed{density-dependent} mark to galaxies in the estimation of the projected marked correlation function. Two gravity models are compared: general relativity (GR) and $f(R)$ gravity. We build mock catalogues which, by construction, reproduce the measured galaxy number density and \textRed{two-point} correlation function of the LRG samples, using the halo occupation distribution model (HOD). A range of HOD models give acceptable fits to the observational constraints\textRed{,} and this uncertainty is fed through to the error \textRed{i}n the predicted marked correlation functions. The uncertainty from the HOD modelling is comparable to the sample variance for the SDSS-III LRG samples. Our analysis shows that current galaxy catalogues are too small for the test to distinguish a popular $f(R)$ model from GR. However, upcoming surveys with a better measured galaxy number density and smaller errors on the \textRed{two-point} correlation function, or a better understanding of galaxy formation, may allow our method to distinguish between viable gravity models. 
\end{abstract}




\section{Introduction} \label{sec:introduction}

After the discovery of the accelerating cosmic expansion, $\Lambda$CDM became the standard  cosmological model \citep{Riess98,Perlmutter1999}. Nevertheless, the cosmological constant in this model remains unappealing from a theoretical perspective, which has motivated efforts to look at gravity models beyond general relativity (GR) to explain the accelerated cosmic expansion \citep{Austin2016}. Recently, theories that modify the model of gravity by adding Lagrangian metric variations of the scalar field have been studied intensively  \citep{Clifton2011}. However, some of these modified gravity (MG) models have been ruled out by the detection of gravitational waves and their optical counterparts with the same propagation speed \citep{Creminelli_Filippo2017,Ezquiaga2017,Baker2017}. Such tight constraints illustrate the way in which a range of modified gravity models remain viable and demonstrate the need to devise new probes of gravity  \citep{Heymans2018,Baker2021,Arai2023}.

A model that is a simple extension of GR is the $f(R)$ model of gravity \citep{DeFelice2010}, in which the Ricci scalar, $R$, is perturbed in the Einstein-Hilbert action by the addition of a function $f(R)$. This modification acts to enhance gravity, by producing an effective `fifth force' that reshapes the distribution of matter over certain scales. However, the $f(R)$ model includes a screening mechanism that hides this new physics on scales where GR works well \citep{Khoury2004}, allowing this model to satisfy solar system constraints. This elusive fifth force has to be searched for on cosmological scales where gravity is the dominant force shaping the formation of large-scale structure. Currently, constraints on the amplitude of the fifth force are obtained from observations of the abundance of massive clusters of galaxies \citep{Cataneo2015}, and weak lensing peak statistics \citep{Liu2016}; modelling forecasts of these probes for next generation surveys have helped to add more constraints on MG models \citep{Liu2021,HD2022}.

This paper is the second in a series about a new test of gravity which uses the marked correlation function. The original idea was proposed by \cite{White:2016}, who suggested using a mark based on the local density of a galaxy to compute the marked correlation function, with the aim of using this to distinguish between gravity models. This idea was applied in simulations of different gravity models by  \cite{Armijo2018} and \cite{Hernandez2018}. 
In Paper I, we introduced a pipeline to apply the marked correlation function as a diagnostic of gravity, in which a halo occupation distribution (HOD) model was used to populate $N$-body simulations of different gravity models with galaxies. A key step in our analysis was the construction of mock catalogues which match the available observational constraints, namely the unweighted clustering of galaxies and their abundance, in all of the gravity models to be tested. This step adds an important contribution to the error budget on the predicted marked correlation function, which as we show later can be comparable to the same variance which results from the volume probed. In Paper II we describe the application of our method to current large-scale galaxy catalogues, discussing the properties of the samples studied in more detail than in Paper I.  

Other studies have investigated using the marked correlation function as a probe of gravity. 
\cite{Satpathy2019} estimated the marked correlation function for SDSS-III BOSS galaxies using the LOWZ sample. These authors found the LOWZ measurements agreed with simulations of GR-$\Lambda$CDM in redshift space on scales between $6 < s / (\textrm{Mpc}\ h^{-1}) < 69$. Their analysis is restricted to these scales due to the challenge of modelling redshift space distortions (though see \citealt{Cuesta2020} and \citealt{Ruan2022} for recent improvements that extend the modelling down to smaller scales). \cite{Armijo2018} showed that the differences between GR and $f(R)$ gravity are stronger on smaller scales $r < 2\ \textrm{Mpc}\ h^{-1}$ in real space, which still needs to be tested. 

The structure of this paper is as follows. We describe the data, the luminous red galaxy (LRG) samples from SDSS-III BOSS DR12, in Section~\ref{sec:data}. Section~\ref{sec:marked} outlines the estimation of the marked correlation function. In Section~\ref{sec:results} we present the measured marked correlation function for the LOWZ and CMASS samples, and discuss how well these results agree with the mock catalogues made from the GR and $f(R)$ simulations, considering the various sources of error. In \textRed{Section~\ref{sec:discussion}} we consider the implications of these results and speculate on how future observations and improvements in modelling could make the constraints on gravity models using this test more competitive. Note that the $f(R)$ gravity model was outlined in Section 2 of Paper I, and the simulations used here, along with the construction of the mock catalogues were described in Section 3 of the same paper. 

\section{Data} \label{sec:data}

We use the LRG samples from the Baryon Oscillation Spectroscopic Survey (BOSS) \citep{Eisenstein2011,Dwason:2013}, which is part of the SDSS-III program twelfth data release (DR12) \citep{Alam2015}. The LRGs are divided into two samples with different photometric selections that yield galaxies that are separated in redshift: LOWZ, which contains LRGs over the redshift range $0.10 < z < 0.43$, and CMASS which predominately targets galaxies in the redshift interval  $0.43<z<0.70$. We decided to use only the NGC region of both the LOWZ and CMASS samples, instead of using the full NGC+SGC areas for practical convenience. As these patches correspond to different areas on the sky, we need to consider them as different surveys, with different photometric properties and potentially different systematic errors. Furthermore, the NGC region covers twice the solid angle of the SGC, and so dominates the pair counts in clustering estimates. To simplify our analysis we decided to use two subsamples extracted from LOWZ and CMASS which are defined in narrow redshift ranges. \textRed{For LOWZ we choose redshifts in the range $0.240 < z < 0.360$ while for CMASS, we limit the selection to redshifts between $0.474 < z < 0.528$}. This allow\textRed{s} us to perform our analysis with two samples with similar volumes, where one of the samples has a larger number density. Also, by restricting the redshift range in this way, the variation in the number density of galaxies across the sample is greatly reduced. The catalogues are fully described in \cite{Reid2016}, where further details of the galaxy selection and the use of the resulting LRG samples for LSS studies are presented.

 \subsection{Galaxy number density}
 
\begin{figure}
    \centering
    \includegraphics[width=\linewidth]{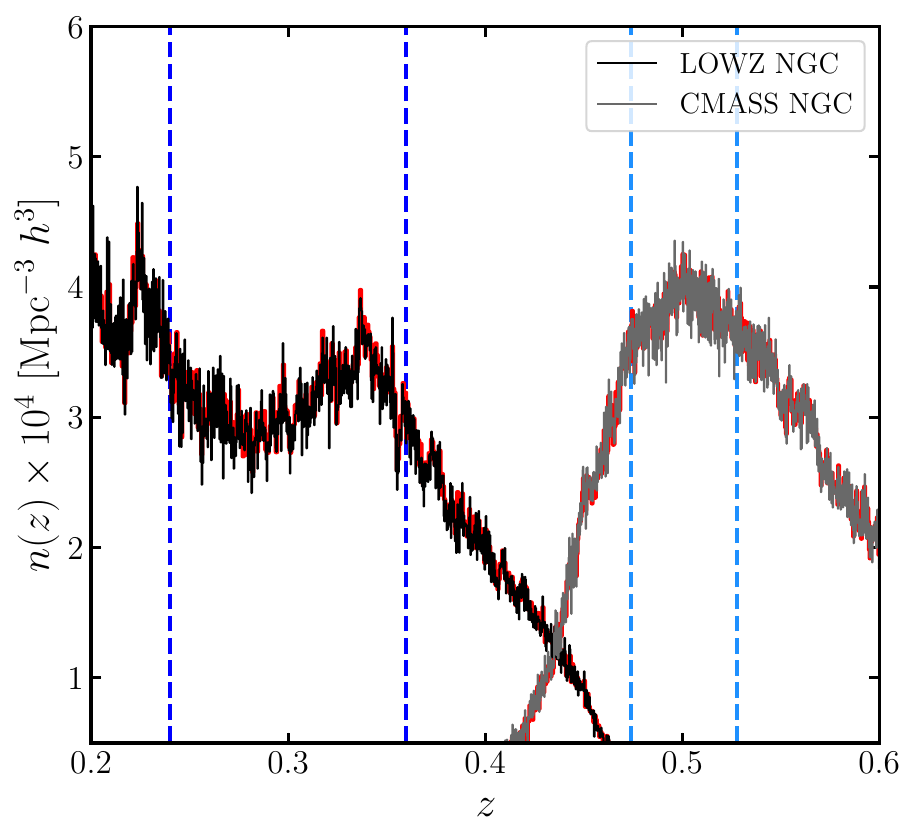}
    \caption{The galaxy number density $n(z)$ as function of redshift $z$ for the BOSS DR12 NGC data. LOWZ (black) and CMASS (gray) samples have different selection functions which lead to different curves for $n(z)$. Over the redshift range shown the number density varies strongly for each sample. We also plot the scaled number density of the random galaxy catalogue (red) from \protect\cite{Reid2016}, used for clustering analyses, and the subsample redshift selection used in  this study LOWZ $0.240 < z < 0.360$ (blue dashed line) and CMASS $0.474 < z < 0.528$ (light blue dashed line).}
    \label{fig:nz_z}
\end{figure}

 As mentioned above, we select narrower redshift range subsamples from the LOWZ and CMASS catalogues to obtain samples for which the number density varies little with redshift, $n(z)$, compared with the full samples. This allows us to treat the data sample as having a constant number density which simplifies the clustering analysis. Fig.~\ref{fig:nz_z} shows the dependence of the LRG number density, $n(z)$, on redshift $z$, after applying the photometric selection in the original LOWZ and CMASS samples. The local variation in $n(z)$ is due to large-scale structure. If we did not restrict the redshift interval studied in this way, we would be introducing new dependencies into the properties (e.g. the weight assigned to each galaxy) that depend on the number density when we compute the marked correlation function. To avoid this problem, we define the number density of the survey to be the number of galaxies divided by the total volume $n_{\textrm {obs}} = N_{\textrm {gal}}/V_s$. By using a more restricted volume for both samples this means that there is less variation in number density, which in turn reduces the error when computing the clustering and marked clustering. The dashed lines in Fig.~\ref{fig:nz_z} show the redshift limits of these new subsamples.
 Using these additional redshift selections results in samples with roughly uniform number densities over the redshift range being considered. 
 We can also compare these new samples with simulations of roughly the same volume when we create the mock catalogues. With these additional redshift selections and the definition of number density given above, the galaxy number density of the LOWZ subsample is $n_{\textrm{g}} = 3.097\times 10^{-4}\ h^{3} \,\textrm{Mpc}^{-3}$, whereas for CMASS the value is 21 per cent higher, $n_{\textrm {g}} = 3.761\times 10^{-4}\ h^{3} \,\textrm{Mpc}^{-3}$. This allows us to evaluate the marked correlation function analysis for samples with different number densities.
 
 \subsection{Galaxy-galaxy two-point correlation function}

 \begin{figure}
    \centering
    \includegraphics[width=\linewidth]{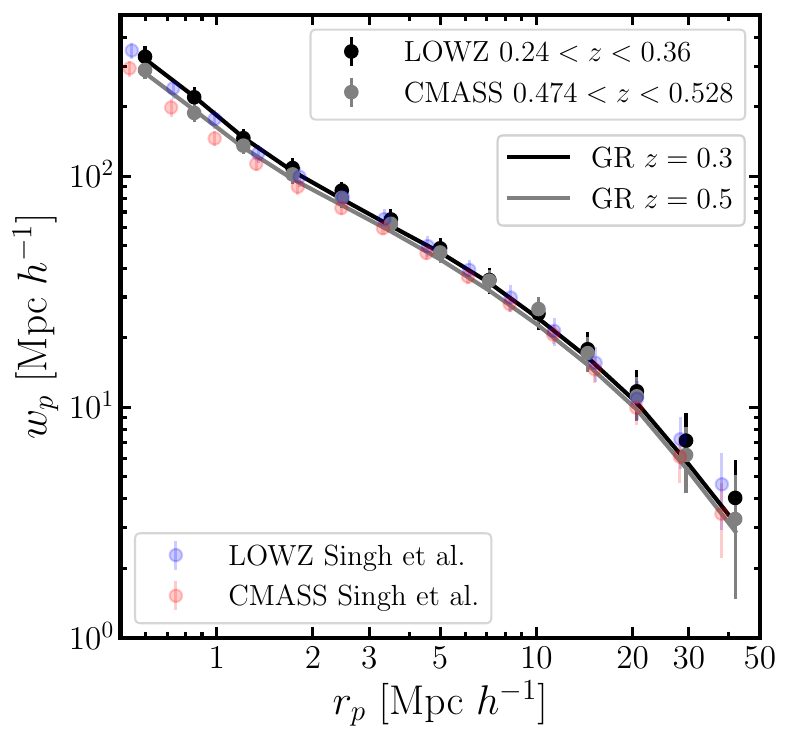}
    \caption{The projected two-point correlation function $w_{\textrm p}$ as a function of the projected perpendicular pair separation $r_{\textrm p}$ for BOSS DR12 NGC. The correlation function is measured from the selected subsamples of LOWZ (black dots) and CMASS (gray dots). Error bars are estimated using jackknife resampling over 100 jackknife regions. Calculations of $w_{\rm p}$ for GR mock catalogues at $z=0.3$ (black line) and $z=0.5$ (gray line) are also shown. We compare our results with those from \protect\cite{Singh2015}, where $w_{\rm p}$ is also calculated for the LOWZ (light bue circles) and CMASS (light red circles) samples over a much wider range of redshifts in each case.}
    \label{fig:2PCF}
\end{figure}
 
 Once we have selected the new restricted redshift range of the subsamples, the next step is to estimate the clustering of galaxies on different scales. The two-point correlation function can be computed as the excess probability of finding a pair of galaxies at a given separation compared with the number of pairs expected in a random distribution of points. Throughout this study, we measure the clustering using the projected correlation function $w_{\textrm{p}}$, which is an integral over the two-point correlation function $\xi(r_\textrm{p},\pi)$, binned in the separation $r_\textrm{p}$ in the projected perpendicular distance, and in the separation parallel to the line-of-sight, $\pi$. The integral of $\xi(r_\textrm{p},\pi)$ is taken over the separation parallel to the line-of-sight direction $\pi$. Clustering measurements as a function of the perpendicular distance $r_\textrm{p}$ can be considered as being in real space (i.e. free from redshift space distortions) in the distant-observer approximation \citep{DavisPeebles1983}. We take this approach instead of using the redshift space two-point correlation function $\xi(s)$ to avoid the influence of small-scale redshift space distortions, which can complicate the prediction of the marked correlation function on such scales. These issues were highlighted  by \cite{Satpathy2019}, in which the marked correlation function of LOWZ is presented in redshift space for pair separations in the range $0.5 < s / ({\rm Mpc h^{-1}}) < 69$. These authors concluded that their results are restricted to these scales by the limited accuracy with which the clustering in redshift space can be modelled on small scales (though for recent improvements in this modelling see \citealt{Cuesta2020} and \citealt{Ruan2022}). To calculate the projected correlation function and obtain the clustering signal in real space we integrate $\xi(r_\textrm{p},\pi)$ in the $\pi$-direction:

\begin{equation}
    \frac{w_\textrm{p}}{r_\textrm{p}} = \frac{2}{r_\textrm{p}}\int_0^{\infty} \xi(r_\textrm{p},\pi){\rm d}\pi. \label{chp3:eq:wp_rp}
\end{equation}

As we are not solving this integral analytically we bin $\xi(r_p,\pi)$ until $\pi_{\rm max}$, which is chosen so that the integral converges to a stable value. Using the correlation function on scales larger than  $\pi_{\rm max}$ tends to add noise to the estimate, depending on the details of the galaxy sample. Considering the range of scales we are interested in, we choose $\pi_{\rm max} = 80h^{-1}$ Mpc, as adopted in \cite{Parejko2013} for the LOWZ data sample. In Fig.~\ref{fig:2PCF} we plot the results for the projected correlation function as a function of the separation perpendicular to the line of sight $r_{\textrm{p}}$ on scales between $0.5 < r_{\textrm{p}} / (h^{-1} \textrm{Mpc}) < 50$ for both the LOWZ and CMASS subsamples. The correlation functions show similar features, with a small offset due to the different number densities that the subsamples have and because the samples probe galaxies with different bias factors at different redshifts. We note that the curves cross one another at $r_{\rm p} = 7\, h^{-1}\, \textrm{Mpc}$, which can be attributed to different slopes being found for the correlation functions of the LOWZ and CMASS galaxies over the range $2 < r_{\textrm{p}} / (h^{-1} \textrm{Mpc}) < 10$. This could be a reflection of the intrinsic differences between LOWZ and CMASS galaxies, with CMASS galaxies having a broader colour selection \citep{Tojeiro2012}. We use the jackknife re-sampling method to compute the uncertainties on the measurements of $w_{\textrm{p}}$ \citep[e.g.][]{Norberg2009}. These calculations can be compared in Fig.~\ref{fig:2PCF} with independent estimates, such as the measurements from \cite{Singh2021}, in which  $w_{\textrm{p}}$ is estimated for the LOWZ and CMASS samples as part of these authors' study of intrinsic alignments. In \cite{Singh2021} $w_{\textrm{p}}$  is calculated using the full redshift ranges of the LOWZ and CMASS samples, with $\pi_{\rm max} = 100h^{-1}\ \textrm{Mpc}$ (see their Fig. 4) . The different set up used in this study in comparison to that used by \cite{Singh2021} can explain the small differences between our results. The broader redshift range used by Singh et al. means a higher volume of the surveyed galaxies, in particular for CMASS (a factor of 6 in volume), which has an impact on the estimation of the uncertainties in $w_{\rm p}$, being approximately a 40\% smaller for their study.

\begin{figure*}
    \centering
    \includegraphics[width=0.48\linewidth]{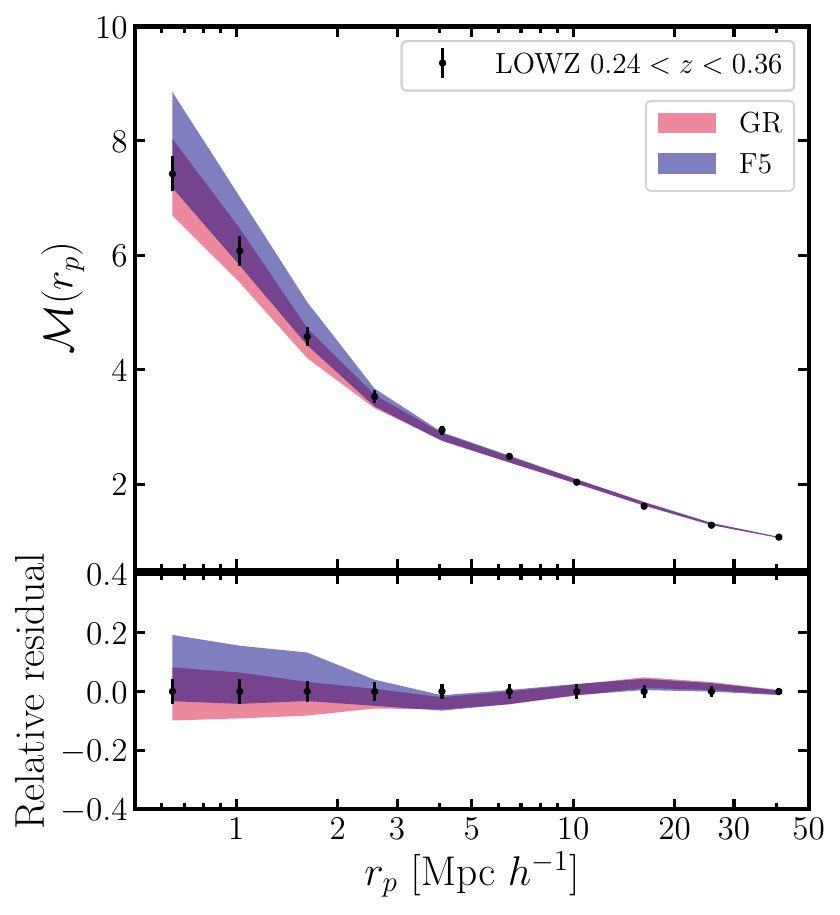}
    \includegraphics[width=0.48\linewidth]{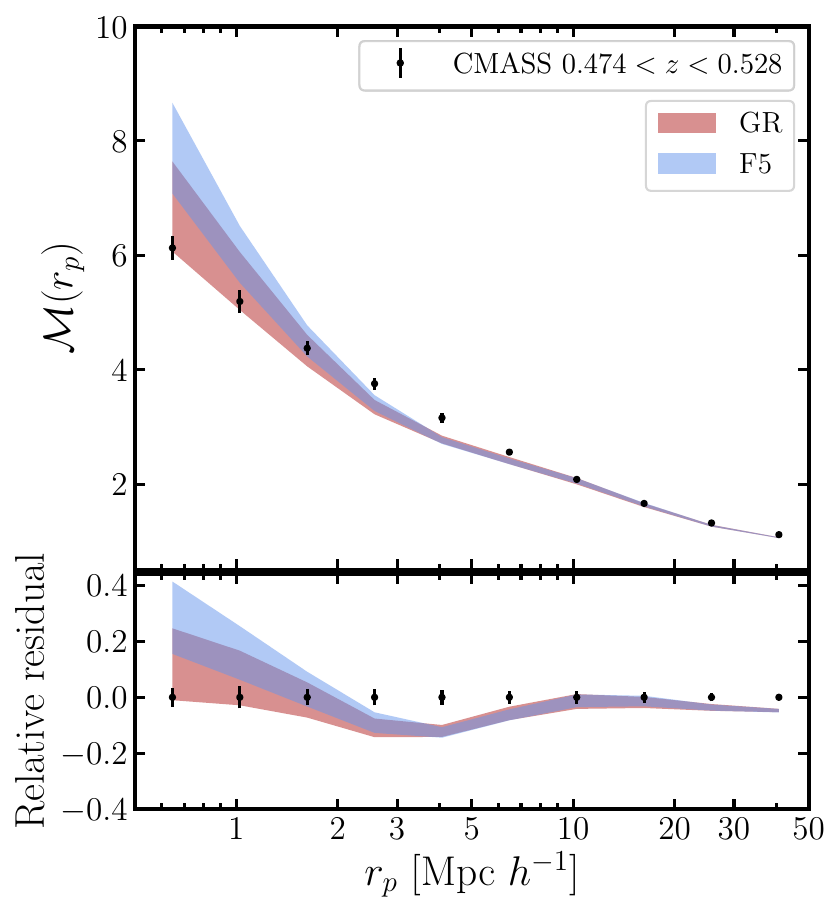}
    \caption{The marked correlation function $\mathcal{M}(r_{\textrm{p}})$ as a function of the projected distance $r_{\textrm{p}}$ for the BOSS galaxy samples and the results from the respective HOD mock galaxy catalogues from the GR (red) and F5 (blue) simulations. Left panel: $\mathcal{M}(r_{\textrm {p}})$ measured from LOWZ (black dots) at $0.24 <z < 0.36$ compared with the HOD mock catalogues within the $1$-$\sigma$ confidence interval from the MCMC fitting of the two-point clustering and number density. Right: same as left panel, but for the CMASS subsample (grey dots) at $0.474 < z < 0.528$. The shaded areas for the models come from selecting the central 68 per cent of all the family of HOD catalogues of each model, GR, F5 at redshift $z=0.3$ (dark red and dark blue) and $z=0.5$ (light red and light blue). The error bars on the data are estimated by applying jackknife resampling to 100 subvolumes of the data. In the bottom panels we show the relative residuals using the data measurements as a reference, meaning that we display $\mathcal{M}^{\textrm{mod}} / \mathcal{M}^{\textrm{data}} - 1$, where  $\mathcal{M}^{\textrm{mod}}$ is the marked correlation function for each set of HOD parameters and $\mathcal{M}^{\textrm{data}}$ is the marked correlation function of LOWZ and CMASS in left and right panels respectively.}
    \label{fig:MCF}
\end{figure*}

\section{Marked correlation function}
\label{sec:marked}
We calculate the marked correlation function of the LOWZ and CMASS samples using marks derived from estimates of the local density. We use the method developed in \cite{Armijo2023}, in which the marked correlation function is estimated in projection (see Section 5 of Paper I). To compute the marked correlation function we use the \textsc{twopcf}\footnote[1]{\url{https://github.com/lstothert/two_pcf}} code to compute $w_{\rm p}(r_{\rm p})$ for the data and mock catalogues; this code supports 
estimators that use weighted pair counts. The code can also efficiently calculate jackknife errors in a {\it single} loop over the galaxy pairs. To compute the marks based on the galaxy's local density we calculate 2D Voronoi tessellations after dividing each sample into several redshift slices. In the case of the LOWZ subsample defined between $0.24 < z < 0.36$, we create 8 redshift slices with a mean thickness of $\Delta \bar{Z} = 38.42\,h^{-1}\, \textrm{Mpc}$, whereas for CMASS, 4 samples are defined with a mean thickness of $\Delta \bar{Z} = 30.72\,h^{-1}\,\textrm{Mpc}$. The projection over $\Delta \bar{Z}$ is the only smoothing applied to the sample, besides the Voronoi tessellation. The slightly smaller slice thickness adopted for the CMASS slices was chosen to preserve $\bar{V}$, the mean volume of a Voronoi cell in each case, the same as in the simulations, due to the higher galaxy number density in the CMASS sample compared to LOWZ. To construct tessellations over the irregular boundary of the survey angular mask, we apply a random sample embedded within a rectangular region covering the survey edges. This results in any holes left by the mask being flagged as very low-density regions during the tessellation step. The only requirement for this random sample wrapping around the survey is that it should oversample the observed $n(z)$ by a large factor. We select this factor to be at least 10 times larger than the $n(z)$ of the galaxies to make sure the result of the marked correlation function converges to stable values. The mark scheme is equivalent to the one presented in \cite{Satpathy2019}, where the marks based on the local density definition are combined with the observational weights when computing the correlation function. We extend the analysis of \textRed{Satpathy et~al}. by making measurements for the CMASS sample as well as for LOWZ.

\textRed{The goodness of fit between the predicted marked correlation function and that measured from the observed samples is quantified in terms of $\chi^2$, defined as}

\textRed{
\begin{equation}
    \chi_{\mu}^2 = (\mathbf{x} - \bm{\mu})^{\intercal} \mathbf{\Sigma^{-1}} (\mathbf{x} - \bm{\mu}),\label{chp5:eq:chi_2_def}
\end{equation}
}
\textRed{where the statistic in question is the marked correlation function $\mathcal{M}$, $\mathbf{x}$ is the realization value of this quantity drawn from the set of parameters, and $\bm{\mu}$ is the observable that we are trying to model. $\mathbf{\Sigma^{-1}}$ is the inverse of the covariance matrix, which includes the uncertainties in the observation of $\bm{\mu}$. The reduced $\chi^{2}$ is obtained by dividing by the number of bins used to estimate the marked correlation function (10 in our case).} 

\section{Results} \label{sec:results}

We plot the measurements of the marked correlation function, $\mathcal{M}(r_{\textrm{p}})$, for the LOWZ and CMASS subsamples in Fig.~\ref{fig:MCF}. We compare these measurements with the predictions for the marked correlation function made using the GR and F5 mock catalogues presented in \textRed{\citep{Armijo2023}}. 
The marked correlation function of the LOWZ sample appears to agree with the predictions from both the GR and F5 models over the range of scales tested. Within the uncertainties introduced by the model, both the GR and F5 results overlap on scales $r_{\textrm{p}} > 3\,h^{-1}\, \textrm{Mpc}$. On smaller scales, the models show a modest difference, but not one that is statistically significant given the LOWZ errors. For the CMASS sample, the results are similar but show somewhat different features: the observational measurements at large projected separations, $r_{\rm p} > 10\,h^{-1}\, \textrm{Mpc}$ are again reproduced by both the GR and F5 models. However, in the CMASS case, there is also a clear mismatch between models and data on scales $2 < r_{\textrm{p}} /(\,h^{-1}\, \textrm{Mpc}) < 10$. For smaller scales, $r_{\textrm{p}} < 2\,h^{-1}\, \textrm{Mpc}$, the data fits the GR model better than F5. Nevertheless, as the model predictions still overlap given the errors, the difference is still marginal. 

The LOWZ data seems to be a slightly better fit to the GR model with $\chi_{\nu,\textrm{GR}}^2 = 1.13$ in comparison to the F5 model which has $\chi_{\nu,\textrm{F5}}^2 = 1.48$, where these reduced $\chi^2$ values are calculated considering the mean of all the valid models shown in Fig.~\ref{fig:MCF}. \textRed{For the CMASS data, these values are $\chi_{\nu,\rm GR}^2 = 6.21$ and $\chi_{\nu,\rm F5}^2 = 14.99$, which are higher in comparison to LOWZ due to the mismatch between the models and data explained above. Hence, these values are not being used to calculate the goodness of fit for the CMASS sample and are not included in the conclusions of this work.}

\subsection{Marked correlation function error analysis}
We now compare the size of different contributions to the uncertainty in the calculation of the marked correlation function. For the data, we resample the catalogues to estimate the sample variance using jackknife errors. To quantify the significance of the mark, we also shuffle the weights for the marked correlation function calculation. In the case of the mocks, in addition to the sources of error listed above, an important contribution to the error estimate comes from the uncertainty in the model used to create the galaxy catalogues, the halo occupation distribution (HOD) model. In Fig.~\ref{fig:errors_HOD_JK}, we compare these sources of uncertainty in units of the marked correlation function in each case. The first uncertainty contribution comes from the sample or cosmic variance, caused by measuring the clustering statistic in a random realization of the underlying cosmology \citep{GilMarin2010}. We use jackknife resampling \citep{Shao1986}, which is a widely used method to estimate the effect of sample variance in clustering studies \citep[e.g.][]{Norberg2009}. The estimation of the jackknife error bar (red line in Fig.~\ref{fig:errors_HOD_JK}) shows a higher fractional uncertainty at small $r_{\textrm{p}}$ than at large separations, which is expected from previous formulations of the marked correlation function \citep{Armijo2018}. Another source of error comes from the correct estimation of weights for individual galaxies, which gives significance to the individual marks when the clustering is computed. This can be estimated by doing a shuffle of the galaxy marks, assigning a random weight to all galaxies, and recomputing the marked correlation function. The random weights will erase any correlation between the marks and the clustering, which will result in $\mathcal{M} = 1$ on all scales. We show the dispersion of 100 shuffling realizations for the mock
in Fig.~\ref{fig:errors_HOD_JK} (blue line). Finally, we also compare with the uncertainty introduced by the HOD modelling when creating the mock data, which is explained in \cite{Armijo2023}. \textRed{The uncertainty estimations in Fig.~\ref{fig:errors_HOD_JK} is divided by the mean of the corresponding marked correlation functions,$\bar{\mathcal{M}}$: this quantity is the mean of the jackknife and shuffling realizations, using a set of HOD parameters with values close to the mean of all the HOD values in our sample}.
This contribution to the error dominates over the others on small scales, which explains the difference in the size of the error bars on the results from the data and the mocks in Fig.~\ref{fig:MCF}. These are the scales on which the marked correlation function has the largest amplitude and hence for which there is the greatest potential to distinguish between different gravity models. Unfortunately, for the LOWZ and CMASS samples we have considered, the error from the range of acceptable HOD models is too large for these datasets to be able to distinguish the F5 gravity model from GR. 

\section{Conclusions and discussion}\label{sec:discussion}
We have applied the marked correlation test of gravity introduced in Armijo et~al. (2023; Paper I) to currently available large-scale structure samples extracted from the LOWZ and CMASS LRG catalogues. We compared these results with predictions made from simulations of the GR and F5 $f(R)$ gravity models, including the uncertainties introduced by the HOD modelling used to populate the simulations with galaxies.

The measurements of the marked correlation function for the LOWZ and CMASS samples show a slight tendency to agree with the GR model better than F5. However, this conclusion is not statistically significant once all sources of error are taken into account. 

In particular, the HOD modelling used to populate $N$-body simulations with galaxies introduces an error that is typically ignored in the assessment of the forecast for a clustering measurement. This error arises because a range of HOD models give acceptable fits to the clustering and galaxy abundance measurements used to constrain the HOD model parameters (see Paper I). 
In Armijo et~al.(2023) we argued that it is essential to fold this HOD model uncertainty through the mock pipeline. Here, we have demonstrated that for the LOW and CMASS samples studied, this contribution to the error budget for the marked correlation function dominates on small scales, compared to sample variance and the error from shuffling the marks. 

\begin{figure}
    \centering
    \includegraphics[width=\linewidth]{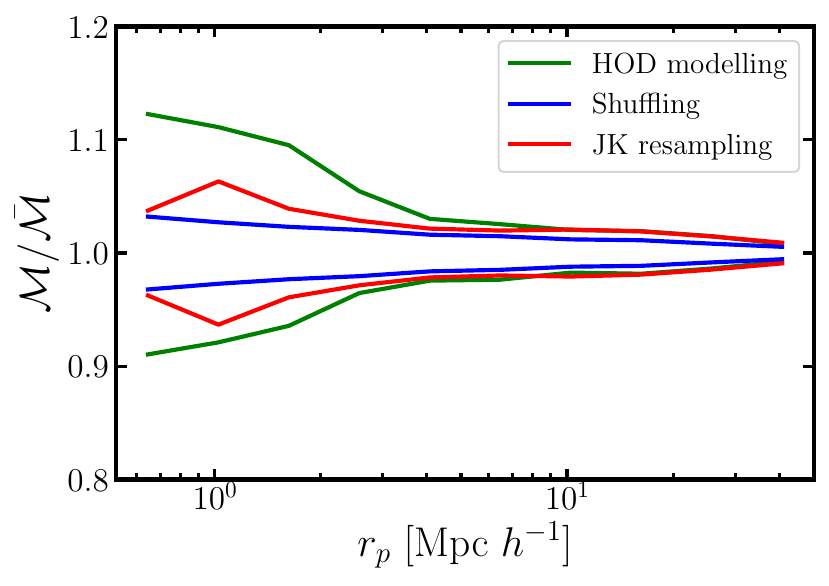}
    \caption{Comparison of the uncertainties estimation of the marked correlation function, $\mathcal{M}$, as function of the scale $r_{\rm p}$ from considering the HOD modelling (green), the jackknife resampling (red) and the effect of shuffling (blue). We use the GR HOD mock catalogues from Armijo et~al. (2023) to calculate $\mathcal{M}(r_{\rm p})$. \textRed{We divide each of the curves by their corresponding mean values $\bar{\mathcal{M}}$ to show the amplitude of the error bars in more detail.}
    }
    \label{fig:errors_HOD_JK}
\end{figure}

When compared to the LOWZ data (left panel in Fig.~\ref{fig:MCF}), the marked correlation is in agreement with both the GR and F5 simulations within the error bars estimated from the HOD modelling. The same analysis is more complex in the case of CMASS data (right panel of Fig.~\ref{fig:MCF}), as there is a disagreement between the proposed models and the data. This disagreement comes from a limitation of the model to replicate the CMASS data, which is comprised of slightly `bluer' galaxies than the ones in the LOWZ sample \citep{Maraston2013}, due to the broader range in both magnitude and colour accepted compared with other LRG samples \citep{Tojeiro2012,Guo2013}; this selection is to increase the number density of galaxies at higher redshift. This selection can be harder to capture with the simple HOD model used here, which could lead to discrepancies between the model and the data. Furthermore, the comparison between the error bars of the model and data in Fig.~\ref{fig:errors_HOD_JK}, indicates that the HOD model introduces more uncertainty (around a factor of 2) on the scales where the disagreement is found. 

We find no sign of any departure from GR for the LOWZ data, which confirms the conclusions reached by \cite{Satpathy2019}, who measured the two-point correlation function in redshift space for separations in the range $6 < s / (\textrm{Mpc}\ h^{-1}) < 69$. Our results are presented in the projected space, extending the calculation down to  small scales with $r_{\textrm{p}} \sim 0.5 h^{-1}\ \textrm{Mpc}$. We can calculate the goodness of fit for the LOWZ data obtaining $\chi_{\nu,\textrm{GR}}^2 = 1.13$ and $\chi_{\nu,\textrm{F5}}^2 = 1.64$, which indicates that LOWZ fits the GR model better. However, the value of $\chi_{\nu,\textrm{F5}}^2$ is not enough to rule out the F5 model with this data alone. For CMASS we note that the higher number density of the sample reduces the estimated error on the uncertainties including sampling variance, which could help to constrain the models further \citep{Seljak2009}. Nevertheless, systematic effects make the data disagree with both models on scales between $2 < r_{\textrm{p}} / (h^{-1} \, \textrm{Mpc}) < 7$ which limits the conclusions we can reach from this dataset. We attribute such differences to the selection function of the CMASS sample, which retains a broader selection of magnitude and colours than the LRG LOWZ sample. This can also be seen in Fig.~\ref{fig:2PCF}, where the projected correlation function of the CMASS sample (grey squares) also behaves differently from the one from LOWZ (black dots). In conclusion, the LOWZ data is consistent with both the GR and F5 simulations. The same conclusion cannot be applied to CMASS, as the marked correlation function is more sensitive to its selection function. 

This leads naturally to speculation about what would need to improve for the test proposed by \textRed{\citep{Armijo2023}} to be in a position to distinguish between currently viable gravity models. The dominant source of error on small scales, on which the marked correlation function is largest, is the allowed range of HOD models. Using a more sophisticated HOD model might improve the performance of the mock at reproducing the clustering measured for the CMASS sample. However, this would come at the expense of greater freedom in a larger HOD parameter space and presumably even greater uncertainty in the marked correlation function on small scales. Alternatively, the HOD model could be replaced by a calculation with less uncertainty, or equivalently, fewer parameters. For example, with a higher resolution $N$-body simulation to hand, a sub-halo abundance matching approach could be used instead, assigning model LRGs to resolved subhalos.

The other way to reduce the uncertainty in the galaxy formation modelling is to improve the measurement of the number density of galaxies, for example by targeting fainter and therefore more abundant galaxies, or by obtaining a better measurement of the two-point correlation function. The latter improvement would be driven by sampling a larger survey volume. This will also have the side effect of potentially reducing the sample variance errors in the marked correlation function, though this is hard to judge without a calculation as the marked clustering is derived from the ratio of correlation functions taken from the same volume. Both of these objectives will be met by upcoming wide-field surveys, such as the DESI survey of LRGs \citep{Zhou2020,Zhou2021}. 




\section*{Acknowledgements}
This work was supported by the World Premier International Research Center Initiative (WPI), MEXT, Japan. JA acknowledges support from CONICYT
PFCHA/DOCTORADO BECAS CHILE/2018 - 72190634.
PN and CMB are supported by the UK Science and Technology Funding Council (STFC) through ST/T000244/1. 
NDP acknowledges support from a RAICES, a RAICES-Federal, and PICT-2021-I-A-00700 grants from the Ministerio de Ciencia, Tecnología e Innovación, Argentina.  We acknowledge financial support from the European
Union’s Horizon 2020 Research and Innovation programme under the
Marie Sklodowska-Curie grant agreement number 734374 - Project
acronym: LACEGAL. 
This work used the DiRAC@Durham facility managed by the Institute for Computational Cosmology on behalf of the STFC DiRAC HPC Facility (www.dirac.ac.uk). The equipment was funded by BEIS capital funding via STFC capital grants ST/K00042X/1, ST/P002293/1, ST/R002371/1 and ST/S002502/1, Durham University and STFC operations grant ST/R000832/1. DiRAC is part of the National e-Infrastructure.

\section*{Data Availability}

The simulations and the generated mock galaxy catalogues used in this paper are available on reasonable request. To use the marked correlation function data, the main author of the papers should be contacted first.



\bibliographystyle{mnras}
\bibliography{example} 






\bsp	
\label{lastpage}
\end{document}